\documentclass{PoS}

\usepackage{dsfont}

\title{Dual representation for massless fermions with chemical potential and U(1) gauge fields}

\ShortTitle{Dual representation for massless fermions}

\author{
\speaker{Christof Gattringer\,}, Thomas Kloiber and Vasily Sazonov\,\thanks{$\,$ This work is supported by the Austrian 
Science Fund FWF, through the DK {\sl Hadrons in Vacuum, Nuclei, and Stars} (FWF DK W1203-N16) and by FWF Grant I 1452-N27. 
We also acknowledge partial support from DFG TR55, {\sl ``Hadron Properties from Lattice QCD''}. We thank Daniel G\"oschl for discussions.}\\ \\
Universit\"at Graz, Institut f\"ur Physik, Universit\"atsplatz 5, 8010 Graz, Austria \\ \\
        E-mail: \email{christof.gattringer@uni-graz.at}, \email{thomas.kloiber@uni-graz.at},\email{vasily.sazonov@uni-graz.at}}

\abstract{Complex action problems coming from either a chemical potential or a topological term have been solved for several models in 
recent years by mapping them to so-called dual degrees of freedom. In terms of these dual variables the partition sum has only 
real and positive contributions such that a Monte Carlo simulation is possible. In this paper we discuss a dual representation
for massless staggered fermions in two dimensions coupled to a U(1) gauge field. We include a topological term, and for the case of several flavors 
with vanishing total charge also
a chemical potential. We show that the real and positive dualization can also be generalized to a system of nanowires (1+1 dimensional fermions) 
coupled to a 3+1 dimensional U(1) gauge field.}

\FullConference{The 33rd International Symposium on Lattice Field Theory\\
		14 -18 July 2015\\
		Kobe International Conference Center, Kobe, Japan}

\begin{document}

\section{Introduction}
A reliable ab-initio lattice simulation of QCD at finite density is currently one of the great challenges for the lattice community. 
The main technical obstacle is the fact that at finite density the action $S$ becomes complex and the Boltzmann factor 
$e^{-S}$ cannot be used as a probability in a stochastic process. This so-called complex action problem has so far 
considerably slowed down the analysis of the QCD phase diagram with lattice methods. In a similar way also the addition
of a topological term gives rise to an imaginary part and hence a complex action problem. 

In recent years several lattice field theories with a chemical potential or a topological term were successfully simulated
using an alternative set of degrees of freedom. The 
key step for success was finding a transformation of the partition sum to so-called dual variables where the partition
sum has only real and positive contributions. The dual variables are loops for matter fields and surfaces for gauge fields  
and a dual Monte Carlo simulation directly updates these degrees of freedom. A discussion of the developments for
overcoming the complex action problem with dual variables can be found in some of the review talks at the lattice conferences 
in recent years \cite{rev1}--\cite{rev6}.

In this contribution we discuss new results that also include fermions, where the Grassmann nature of the fields gives
rise to additional minus signs. More specifically we present a real and positive dual representation for massless staggered fermions
in two dimensions coupled to U(1) gauge fields. We show that both complex action problems, the one from a chemical potential, as well 
as the one from a topological term are overcome by our dualization \cite{2dfermions}. 
Furthermore we show that the dualization can be generalized to
the case of 1+1 dimensional fermions coupling to a 3+1 dimensional U(1) gauge field, i.e., a system of nanowires interacting
with the electromagnetic field. 

\section{Dual representation for 2-d staggered fermions coupled to a U(1) gauge field}

Let us begin with first introducing the 2-dimensional theory and discussing its dualization. We consider a partition sum of the form
\begin{equation}
\label{eq_conv_part}
Z  \; = \; \int \! \mathcal{D}[U]\mathcal{D}\big[\,\overline{\psi},\psi \,\big]~e^{-\, S_G[U]  \, -i\,\theta \, Q[U] \, - \, S_\psi[U,\overline{\psi},\psi] }  \; .
\end{equation}
The gauge fields are described by U(1)-valued link variables $U_\nu(n) = \exp(i A_\nu(n)), \, A_\nu(n) \in [-\pi, \pi]$ living on the
links $(n,\nu)$ of a 2-dimensional lattice with periodic boundary conditions. The fermions are one-component Grassmann valued 
fields $\psi(n)$ and $\overline{\psi}(n)$ living
on the sites $n$ of the lattice. They obey boundary conditions that are periodic in space (the 1-direction) and anti-periodic in time
(the 2-direction). $V = N_1 N_2$ denotes the total number of lattice sites.
For the case of finite chemical potential we need overall neutrality of the system due to Gauss law, which implies that 
in that case we have several flavors of fermions such that the sum of all their charges vanishes. The path integral measures 
are the product of Haar measures for the gauge fields on all links and the product of Grassmann measures for all fermionic degrees of freedom.

We use the Wilson gauge action (the constant term was dropped for simplicity) and a simple field theoretical definition 
of the topological charge (compare \cite{topo2d}),
 
\begin{equation}
\label{eq_conv_gauge}
S_G[U] \, + \; i\,\theta \, Q[U]  \; = \; -\beta \sum_{n} \mbox{Re} \, U_p(n)  + i \theta \frac{1}{2 \pi} \sum_n \mbox{Im} \, U_p(n) 
 \; = \; - \sum_{n} \big[ \, \eta U_p(n) \, + \, \overline{\eta} U_p(n)^{-1} \, \big] \; ,
\end{equation}
where we have introduced the abbreviations
$\eta = \frac{\beta}{2}-\frac{\theta}{4\pi}$, $\overline{\eta} = \frac{\beta}{2}+\frac{\theta}{4\pi}$. By $U_p(n)$ we denote the
plaquettes $U_p(n) \, = \,U_1(n) \, U_2(n+\hat{1}) \, U_1 (n+\hat{2})^{-1} \, U_2(n)^{-1}$. The action for 
massless staggered fermions with a chemical potential $\mu$ is given by 
\begin{equation}
\label{psiaction}
S_\psi[U,\overline{\psi},\psi] \; = \;  
\sum_{n, \nu } \frac{\gamma_\nu(n)}{2} \Big[ e^{\mu \delta_{\nu,2}}  \, 
U_\nu(n) \, \overline{\psi}(n) \, \psi(n +\hat{\nu}) \; - \; 
e^{- \mu \delta_{\nu,2}} \, U_\nu(n)^{-1} \, \overline{\psi}(n + \hat{\nu}) \, \psi(n) \Big] \; ,
\end{equation}
where the staggered sign function $\gamma_\nu(n)$ is defined as
$\gamma_1(n)  = 1$ and $\gamma_2(n) \; = \; (-1)^{n_1}$. The chemical potential couples only to the hopping terms in time direction (the 2-direction)
and gives a different weight to forward and backward temporal hopping.

We begin the dualization by first integrating out the fermions in the fermionic path integral $Z_\psi[U]$ 
for one flavor in a background gauge field, $Z_\psi[U]
= \int \! \mathcal{D}\big[\,\overline{\psi},\psi \,\big]~e^{- \, S_\psi[U,\overline{\psi},\psi] }$. The Grassmann integrals at a site $n$ are 
non-zero only if the integrand contains exactly the factor $\psi(n) \overline{\psi}(n)$.  These factors have to be generated by expanding
the exponential $e^{- \, S_\psi[U,\overline{\psi},\psi] }$. Such an expansion provides the hopping terms $\overline{\psi}(n) \, \psi(n +\hat{\nu})$ and 
$\overline{\psi}(n + \hat{\nu}) \, \psi(n)$ located on the links of the lattice. The hopping terms have to be combined such that all 
$\psi(n)$ and $\overline{\psi}(n)$ appear exactly once. Two types of contributions are possible which we illustrate in Fig.~\ref{fig1} below:  
A forward and the corresponding backward hop are combined into a so-called dimer (links with double lines in Fig.~\ref{fig1}). The second
possibility are closed oriented loops made from the hopping terms. Note that for a complete saturation of the Grassmann integral 
each site of the lattice has to be either run through by a loop or be the endpoint of a dimer (compare Fig.\ref{fig1}). 

The Grassmann variables need to be reordered for the integration, a step 
which generates signs for each contribution. Further signs come from the staggered sign function $\gamma_\nu(n)$, from the relative sign
between forward and backward hopping, and from the anti-periodic temporal boundary conditions. For the dimers all signs cancel. 
For a loop $l$ the situation is more involved: There is an overall sign for each loop coming from reordering the Grassmann variables. 
Another sign comes for each winding of the loop around compactified time because of the temporal boundary conditions, i.e., 
we pick up a factor $(-1)^{W(l)}$ where $W(l)$ is the winding number of the loop $l$ around compactified time. 
The relative minus sign between forward and backward hopping gives rise to $(-1)^{\frac{L(l)}{2}}$, where $L(l)$ is the 
length of the loop and we assume that both the spatial and the temporal extent of the lattice are multiples of 4, such that also winding 
loops are taken into account in this simple notation. 
Finally the contributions from the staggered sign functions combine into $(-1)^{P(l)}$, where  
$P(l)$ is the number of plaquettes inside the loop. This factor is not a-priori obvious, but can be proven by 
recursively building up admissible loops \cite{2dfermions}. Thus the overall sign for an admissible configuration of loops and dimers is given by
$(-1)^{N_L + \sum_l [ \frac{L(l)}{2} + W(l) + P(l) ]}$, where $N_L$ denotes the total number of loops in the configuration, and the sum runs over
all these loops $l$. Note that not all loops are admissible since the interior of a loop has to be such that it can 
be filled with dimers. For massive fermions this restriction does not apply, since the mass term allows for 
monomers $\overline{\psi}(n) \psi(n)$ which saturate the Grassmann integral at a single site. In that case additional sign factors appear.

When expanding the fermion action, each hopping term comes with the corresponding link variable $U_\nu(n)$ (or $U_\nu(n)^{-1}$ 
for backward hopping). For dimers the link variables cancel. The loops, however, are dressed with the gauge variables 
on the links of their contour. Furthermore, the temporal links also come with factors $e^{\pm \mu}$ for forward and backward hopping. 
For dimers and loops that close trivially these factors cancel. Only for loops $l$ with a non-zero winding number $W(l)$ 
around the temporal direction there remains a non-trivial factor $e^{\mu N_2 W(l)}$. Finally the factors $1/2$ give rise to an irrelevant overall
factor $(1/2)^V$ for the partition sum.

Thus we can now summarize the result for the fermionic partition function $Z_\psi[U]$. It is given as a sum over  
the set $\{l,d\}$ of all admissible configurations of loops and dimers,
\begin{equation}
Z_\psi[U] \; = \; \left( \frac{1}{2} \right)^{V} \!\! \sum_{\{l,d\}}  \;  (-1)^{N_L + \sum_l \big[ \frac{L(l)}{2} + W(l) + P(l) \big]} \; 
e^{\, \mu \, N_2 \sum_l  W(l)}
\; \prod_l \! \prod_{(n,\nu)\in l} U_\nu(n)^{s_\nu(n)} \; .
\label{zfdual}
\end{equation}
The loops are oriented and dressed with the gauge link variables $U_\nu(n)$ along their contour and the exponents $s_\nu(n)$ are 
$+1$ for links $(n,\nu)$ run through with positive orientation and $s_\nu(n) = -1$ for negative orientation.

\begin{figure}[t!]
\begin{center} 
\includegraphics[width=7cm,type=pdf,ext=.pdf,read=.pdf]{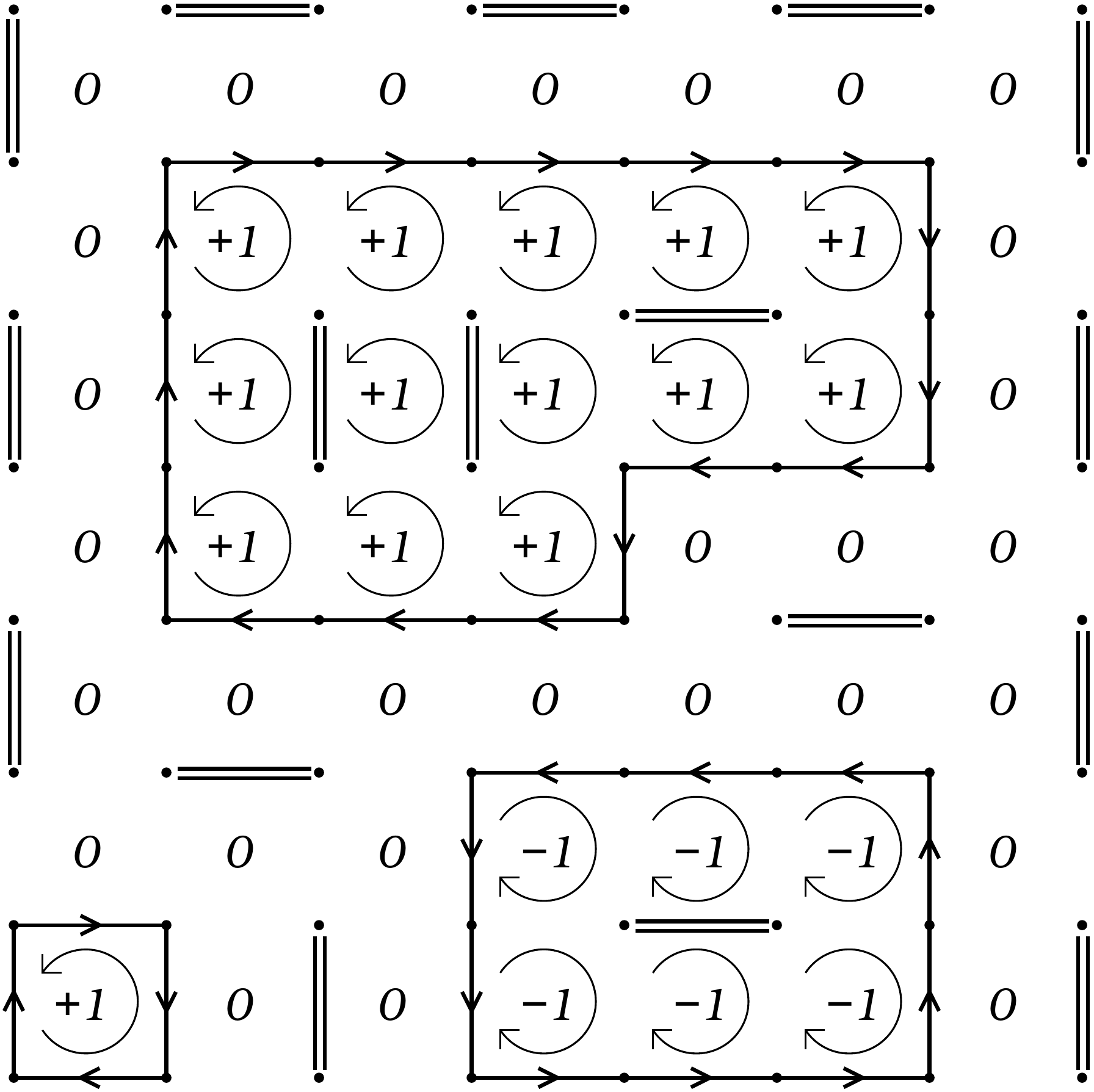}
\end{center}
\caption{An admissible configuration on an $8 \times 8$ lattice. The fermions give rise to oriented loops and dimers on the links of the lattice.
The link variables along the contour of the loop must be compensated by oriented 
plaquettes from the expansion of the gauge action.
The corresponding values of the plaquette occupation numbers are written in the plaquettes (0 for plaquettes outside of the loops).}
\label{fig1}
\end{figure}

The next step in the dualization of the model is to integrate out the fermions. For that purpose we write the Boltzmann factor 
$e^{-S_G[U] - i \theta Q[U]}$  as a product of factors
\begin{equation}
e^{ \, \eta \, U_p(n) \, + \, \overline\eta \, U_p(n)^{-1}}  \; = \; \sum_{p(n) \in \mathds{Z}}  \, 
I_{p(n)}\!\left(2 \sqrt{\eta \overline{\eta}} \, \right) \, \left( \sqrt{\frac{\eta}{\overline{\eta}}} \, \right)^{p(n)} \; U_p(n)^{ \,p(n)} \; ,
\label{gaugebessel}
\end{equation}
which are related to the generating functional of the generalized Bessel functions $I_p$. This relation gives rise to the
power series on the rhs.~of Eq.~(\ref{gaugebessel}). We refer to the expansion indices $p(n)$ attached to the plaquettes
as the plaquette occupation numbers.

In the form on the rhs.\ of (\ref{gaugebessel}) the powers of
the plaquettes $U_p(n)$ can now be used to cancel the link variables $U_\nu(n)$ along the contours of the fermion loops. 
This cancellation is necessary, due to the identity $\int dU_\nu(n) \, U_\nu(n)^{k} \; = \; \delta_{k,0}$ for the integration over the
U(1) valued link variables $U_\nu(n)$. It is straightforward to see that this identity implies that we need to fill the interior of the fermion loops 
with plaquettes as illustrated in Fig.~\ref{fig1}. The values of the plaquette occupation numbers $p(n)$ needed for filling the loops then
determine the weight $\prod_n I_{p(n)}\!\left(2 \sqrt{\eta \overline{\eta}} \, \right) \, \left( \sqrt{\frac{\eta}{\overline{\eta}}} \, \right)^{p(n)}$
of a configuration. We remark that the condition of filling loops with plaquettes 
implies that the total winding number of all loops $\sum_l W(l)$ is even such that after integrating the gauge fields 
the winding number drops out from the sign factors. This also holds for the case of several flavors.

The last step for establishing the real and positive dualization is to show that the remaining sign factors
$(-1)^{N_L + \sum_l [ \frac{L(l)}{2} + P(l) ]}$ vanish. The corresponding proof given in \cite{2dfermions} is based on the observation that the
overall sign can be factorized into the product of the signs for the individual loops of a configuration. These individual loops can then be
built up recursively with three steps, and for each step one can monitor the change of the characteristics $L(l)$ and $P(l)$. In this way one 
can show that $N_L + \sum_l [ \frac{L(l)}{2} + P(l) ]$ is an even number such that all minus signs vanish. For the case of several flavors, this
result holds individually for the loops of each flavor. 

The final result for the dual representation of the partition sum with $N$ flavors reads (we have dropped the irrelevant overall factor)

\begin{equation}
Z \;\; =   \sum_{\{l_1\, .. \, l_N,d_1 \, .. \, d_N,p\}}  
\prod_n I_{p(n)}\!\left( 2 \sqrt{\eta \overline{\eta}} \, \right) \, \left( \sqrt{\frac{\eta}{\overline{\eta}}} \, \right)^{p(n)} \, 
\prod_{j=1}^N e^{\, \mu_j \, N_2 \sum_{\,l_j}  W(l_j)}\; .
\label{zfinal2}
\end{equation}
Here the sum runs over all configurations of loops $l_j$ and dimers $d_j$ for all $N$ flavors $j = 1, 2, \, ... \, N$, and each site of the lattice
has to be run through by a loop or be the endpoint of a dimer for each flavor. Furthermore the configurations have to be such that at each link 
the total flux from all matter loops of different flavor running through that link can be compensated by plaquettes. 
Here the counting is such that each flavor contributes to the flux according to the charge it carries. We have introduced chemical 
potentials $\mu_j, \, j = 1,2, \, ... \, N$ for all flavors. Obviously all contributions to the partition sum are real and positive for arbitrary values
of the $\mu_j$ and also for finite vacuum angle $\theta$, which enters in  
$\eta = \frac{\beta}{2}-\frac{\theta}{4\pi}$, $\overline{\eta} = \frac{\beta}{2}+\frac{\theta}{4\pi}$. Thus the complex action problem is solved.

The simplest possible observables are obtained as derivatives of $\ln Z$ with respect to the parameters $\beta$, $\theta$ and $\mu_j$. After 
suitable normalization this  gives rise to the expectation values of the plaquette $\langle U_p \rangle$, 
the topological charge $\langle Q \rangle$, and the particle number
densities $\langle n_j \rangle , \, j = 1,2 \, ... \, N$ for all flavors. Second derivatives provide the corresponding susceptibilities.

\section{Generalization to 2-d fermions coupled to a 4-d U(1) gauge field}

It is interesting to note, that the proof for the existence of a real and positive dual representation can be generalized to a system of nanowires
interacting with 3+1 dimensional U(1) gauge fields. While the dualization of 3+1 dimensional U(1) gauge theory and 3+1 dimensional 
abelian gauge Higgs models is well understood 
\cite{U1,surfaceworm}, for full 3+1 dimensional lattice QED there are only results for a partial dualization \cite{QED4} and some subsets of matter loops
still have non-trivial signs. However, when the fermions are restricted to a 1+1 dimensional wire embedded in 4 dimensions, then the algebra for
the signs from integrating out the fermions discussed in the last section carries over and we find a real and positive dual representation. 

More specifically the fermion action is given by
\begin{equation}
S_\psi  =  \sum_n \sum_{\nu=1}^4  \sum_{j=1}^N  \theta^{(j)}_\nu(n) \frac{\gamma_\nu(n)}{2} 
\Big[ e^{\mu_j \delta_{\nu,4}}  \, 
U_\nu(n) \, \overline{\psi}_j(n) \, \psi_j(n\! +\!\hat{\nu}) \, - \, 
e^{- \mu_j \delta_{\nu,4}} \, U_\nu(n)^{-1} \, \overline{\psi}_j(n\!+\! \hat{\nu}) \, \psi_j(n) \Big] .
\label{wireaction}
\end{equation} 
Here we have a sum $j = 1,2 \, ... \, N$ over all $N$ wires and by $\psi_j(n)$,  $\overline{\psi}_j(n)$ we denote the 
fermionic degrees of freedom on the wires. The wires 
are defined to be straight lines along one of the three spatial directions ($\nu = 1,2,3$). By $\theta^{(j)}_\nu(n)$ we denote a step function which is
1 for all links that are relevant for the propagation of the fermions on wire number $j$ and 0 for all other hops. These are 
the spatial links on the wire and 
all related temporal hops ($\nu = 4$), i.e., the $\theta^{(j)}_\nu(n)$ activate the space-time plane for the wire number $j$. 
In the path integral measure only these fermion degrees of freedom are integrated over. 
We allow for a different chemical potential $\mu_j$ for each wire which is coupled to the corresponding temporal hops. Finally, here 
$\gamma_\nu(n)$ now is the 4 dimensional staggered factor $\gamma_1(n) = 1$, $\gamma_2(n) = (-1)^{n_1}$, $\gamma_3(n) = (-1)^{n_1 + n_2}$,
$\gamma_4(n) = (-1)^{n_1 + n_2 + n_3}$. 

The fermionic path integral $Z_\psi[U]$ that corresponds to (\ref{wireaction}) can be calculated in the same way as for the 
2-d system discussed in the previous section. The calculation of the sign factors for the fermion loops proceeds in exactly the same way, since
the 4-dimensional $\gamma_\nu(n)$ have the same algebra as in the 2-d case if the propagation is restricted to the planes defined by the
functions $\theta^{(j)}_\nu(n)$. Thus the fermionic partition function is simply a product of 2-d partition functions of the form (\ref{zfdual}). 
For each wire $j$ the corresponding loops $l_j$ and dimers $d_j$ live on the corresponding plane defined by $\theta^{(j)}_\nu(n)$. Note, however,
that these planes are now embedded in 4 dimensions and the gauge variables along the loops $l_j$ are fully 4 dimensional. 

The link variables still have to be integrated over in the path integral. Again we need to compensate the gauge fields dressing the loops
with plaquettes from expanding the Boltzmann factor for the gauge action. As in 2-d we use the Wilson gauge action 
which now is a sum over all plaquettes embedded in 4 dimensions,
$S_G[U] = -\beta \sum_n \sum_{\tau < \sigma} \mbox{Re} \, U_\tau (n) U_\sigma(n\!+\!\hat{\tau}) U_\tau(n\!+\!\hat{\sigma})^* U_\sigma(n)^*$. 
Similar to the 2-dimensional case we can write the Boltzmann factor as a product over all plaquettes and use the expansion formula
(\ref{gaugebessel}) with $\eta = \overline{\eta} = \beta$. 
Now we have plaquette occupation numbers $p_{\tau\sigma}(n) \in \mathds{Z}$ for plaquettes 
in 4 dimensions. The plaquette occupation numbers have to be chosen such that all gauge variables along the 
loops are compensated and the corresponding weight factors are again given by the modified Bessel functions $I_p$. 
As before we allow for different charges in the wires and the total charge on a link is the sum of the contributions from all loops visiting that link.

The final form of the dual partition sum for the system of nanowires now is given by
\begin{equation}
Z \;\; =  \sum_{\{l_1\, .. \, l_N,d_1 \, .. \, d_N,p\}}  
\prod_{n,\tau < \sigma} I_{p_{\tau,\sigma}(n)}\!\left( 2 \beta \right) \,   
\prod_{j=1}^N e^{\, \mu_j \, N_4 \sum_{\,l_j}  W(l_j)}\; .
\label{zfinal2}
\end{equation}
The sum runs over all loop and dimer configurations for all wires $j = 1,2 \, ... \, N$. The loops $l_j$ and dimers $d_j$ for a wire $j$ 
have to fill the plane corresponding to the wire in an admissible way as specified in the previous section. The total 
charge fluxes along all links loops have to be compensated by the plaquette occupation numbers $p_{\tau, \sigma}$
and the weight factor is given by the product over the corresponding modified Bessel functions. Finally, the chemical potentials
$\mu_j$ again enter via the temporal winding numbers $W(l_j)$ of the corresponding loops. As in the 2-d case observables 
are obtained as derivatives of $\ln Z$ with respect to $\beta$ and the $\mu_j$.

\newpage
\section{Concluding remarks}
In this contribution we have presented new results for the program of identifying dual representations for lattice field theories with a 
complex action problem. More specifically we describe real and positive dual representations for 1+1 dimensional massless staggered
fermions interacting with U(1) gauge fields in 1+1 or 3+1 dimensions. Both complex action problems, from either finite chemical potential 
or from a topological term are overcome in the dual formulation. 

It is interesting to note that the positivity of the dual representation is spoiled if the fermions become massive. This is somewhat unexpected 
since in general the complex action problem is expected to become milder when the fermion mass is increased. We interpret this finding as
a manifestation of the well known fact that in 2 dimensions massless fermions can be bosonized also in the continuum. A numerical 
investigation of the 2-dimensional system with a finite vacuum angle $\theta$ and non-zero chemical potential is currently in preparation.

\end{document}